\documentclass{article}
\usepackage{dcase2021,amsmath,graphicx,url,times,booktabs,amsfonts,tabularx,siunitx, microtype}
\usepackage[numbers]{natbib}

\title{AUDIO CAPTIONING TRANSFORMER}

\name{Xinhao Mei$^{1}$,
      Xubo Liu$^{1}$,
      Qiushi Huang$^{2}$, 
      Mark D. Plumbley$^{1}$,
      Wenwu Wang$^{1}$}
\address{$^1$ Centre for Vision, Speech and Signal Processing (CVSSP),\\
        $^2$ Department of Computer Science,\\
        University of Surrey, UK
        }

\begin{document}

\ninept
\maketitle

\begin{sloppy}

\begin{abstract}
Audio captioning aims to automatically generate a natural language description of an audio clip. Most captioning models follow an encoder-decoder architecture, where the decoder predicts words based on the audio features extracted by the encoder. Convolutional neural networks (CNNs) and recurrent neural networks (RNNs) are often used as the audio encoder. However, CNNs can be limited in modelling temporal relationships among the time frames in an audio signal, while RNNs can be limited in modelling the long-range dependencies among the time frames. In this paper, we propose an Audio Captioning Transformer (ACT), which is a full Transformer network based on an encoder-decoder architecture and is totally convolution-free. The proposed method has a better ability to model the global information within an audio signal as well as capture temporal relationships between audio events. We evaluate our model on AudioCaps, which is the largest audio captioning dataset publicly available. Our model shows competitive performance compared to other state-of-the-art approaches.
\end{abstract}


\section{Introduction}
\label{sec:intro}
Automated audio captioning (AAC) is concerned with describing an audio clip using natural language and is a cross-modal translation task at the intersection of audio processing and natural language processing. Generating a meaningful description for an audio clip not only needs to determine what audio events are presented, but also needs to capture and express their spatial-temporal relationships. Audio captioning is practically useful in applications such as assisting the hearing-impaired to understand environmental sounds, retrieving multimedia content, and analyzing sounds for security surveillance. 

Unlike image and video captioning, which have been studied in computer vision (CV) for a longer time, audio captioning is a task investigated only recently \cite{drossos2017automated}. With the announcement of the AAC task in DCASE 2020 and 2021, this topic has attracted increasing attention, and several methods have been proposed \cite{wu2019audio, kim2019audiocaps, ikawa2019neural}. The AAC task is usually treated as a sequence-to-sequence problem, and existing methods are typically based on an encoder-decoder architecture, where the decoder generates words according to the audio features extracted by the encoder. Early works often adopted an ``RNN-RNN'' architecture with an attention mechanism \cite{drossos2017automated, kim2019audiocaps}. However, RNNs can be limited in modeling long-term temporal dependencies in an audio signal. Recently, CNNs have become a dominant approach in audio-related tasks (audio tagging and sound event detection) \cite{kong2020panns}, with many researchers 
using pre-trained CNNs as the audio encoder,  which significantly improved the performance in these systems \cite{xu2021investigating}. More recently, inspired by the great success of the Transformer model in natural language processing \cite{vaswani2017attention}, the RNN decoder has been replaced by a Transformer decoder in captioning models, and the ``CNN+Transformer'' architecture has been shown to achieve state-of-the-art performance in this area \cite{chen2020audio, xinhao2021_t6}. 

Description of an audio signal needs to capture temporal-spatial relationships between audio objects that may be far apart in time. However, convolution is a local operator and has limitations in modelling temporal information, especially with a long audio signal. This can be alleviated by enlarging receptive fields with deeper convolutional layers. However, such deep CNNs can be hard to train and can lead to over-fitting. To address this problem, we propose an Audio Captioning Transformer (ACT), a convolution-free Transformer network based on the self-attention mechanism. We use log mel-spectrograms as input and split the mel-spectrograms into smaller non-overlapping patches along the time axis. By adopting the self-attention mechanism, each patch can attend to all the other patches at each layer of the encoder, which can model global long-range dependencies among the small mel-spectrogram patches from the beginning. Without the need for down-sampling, the features extracted by Transformer are fine-grained, which can contain detailed local audio topics. 

The Transformer usually requires more training data than CNNs \cite{dosovitskiy2020image}. However, the amount of data currently available for audio captioning is relatively small. To address this issue, the ACT encoder is firstly pre-trained on AudioSet dataset \cite{audioset} as an audio tagging task in order to improve its generalization ability. A class token designed to model the global information of an audio clip is appended at the beginning of each patch sequence and is used to output audio tagging results. As a result, when generating words, the decoder can attend to local and global information of an audio clip simultaneously. The proposed ACT model is evaluated on the AudioCaps dataset \cite{kim2019audiocaps} and shows competitive performance as compared to other state-of-the-art methods. 

The remaining sections of this paper are organised as follows. In Section \ref{sec:related_works}, we introduce the related work. The proposed model is described in detail in Section \ref{sec:method}. Experimental settings are shown in Section \ref{sec:exp}. Results are discussed in Section \ref{sec:results}. Finally, we conclude our work in Section \ref{sec:conclusion}.


\section{Related work}
\label{sec:related_works}
Previous work proposed in audio captioning has been based on deep learning methods with an encoder-decoder architecture. \citet{drossos2017automated} proposed the first approach to AAC using an RNN-based encoder-decoder architecture with an alignment model in between. To control the information contained in the output text, \citet{ikawa2019neural} introduced a conditional parameter called ``specificity'' to guide the caption generation. With the release of two freely available datasets AudioCaps \cite{kim2019audiocaps} and Clotho \cite{drossos2020clotho}, AAC has attracted increasing attention and more approaches have been proposed. \citet{kim2019audiocaps} proposed a model with a top-down multi-scale encoder and aligned semantic attention, which enabled the joint use of multi-level features and semantic attributes. As CNNs have achieved state-of-the-art performance in audio tagging and sound event detection tasks \cite{kong2020panns}, some researchers replaced the RNN encoder with CNNs, which brings significant performance gains \cite{chen2020audio, xu2021investigating}. Recently, Transformer has been introduced as the language decoder with a powerful ability in natural language  generation tasks \cite{chen2020audio, tran2020wavetransformer, koizumi2020transformer}. \citet{takeuchi2020effects} formulated audio captioning as a multi-task learning problem, where they proposed keywords estimation and sentence length estimation to avoid the indeterminacy of word selection. \citet{koizumi2020audio} utilized a pre-trained large-scale language model GPT-2 \cite{radford2019language} with audio-based similar caption retrieval to guide the caption generation. Reinforcement learning was used to optimize the audio captioning models with non-differentiable evaluation metrics \cite{xu2020crnn}.


The Transformer was originally proposed for machine translation and has now become the dominant approach in natural language processing tasks  \cite{vaswani2017attention}. Recently, many researchers adopted the Transformer for computer vision tasks which was shown to approach or outperform the state-of-the-art CNNs-based systems in image recognition. \citet{dosovitskiy2020image} proposed a Vision Transformer (ViT) which was based purely on the attention mechanism, i.e. without using convolution kernels, and applied directly to sequences of image patches for the image classification task. However, a large amount of data are required for pre-training the Transformer models, which limits their adoption. To address this problem, \citet{touvron2021training} introduced Data-efficient image Transformers (DeiT) using a data efficiency training and distillation strategy. Based on ViT and DeiT, \citet{liu2021cptr} proposed a CaPtion TransformeR (CPTR) for image captioning. As the Transformer is designed to deal with sequential data, we argue that the Transformer can be adapted for audio signals, and the self-attention mechanism makes it more suitable to capture temporal relationships between audio features and to model the global information. Inspired by these ViT-related works, we propose the Audio Captioning Transformer (ACT) for audio captioning, which, to our knowledge, has not been done in the literature.

\section{Proposed method}
\label{sec:method}
Fig.~\ref{fig:system_overview} shows the proposed Audio Captioning Transformer model, which is based on the traditional sequence-to-sequence architecture and is convolution-free. The model takes the log mel-spectrogram of an audio clip as input and outputs the posterior probabilities of the predicted words. 

\begin{figure}[!t]
  \centering
  \includegraphics[width=\linewidth]{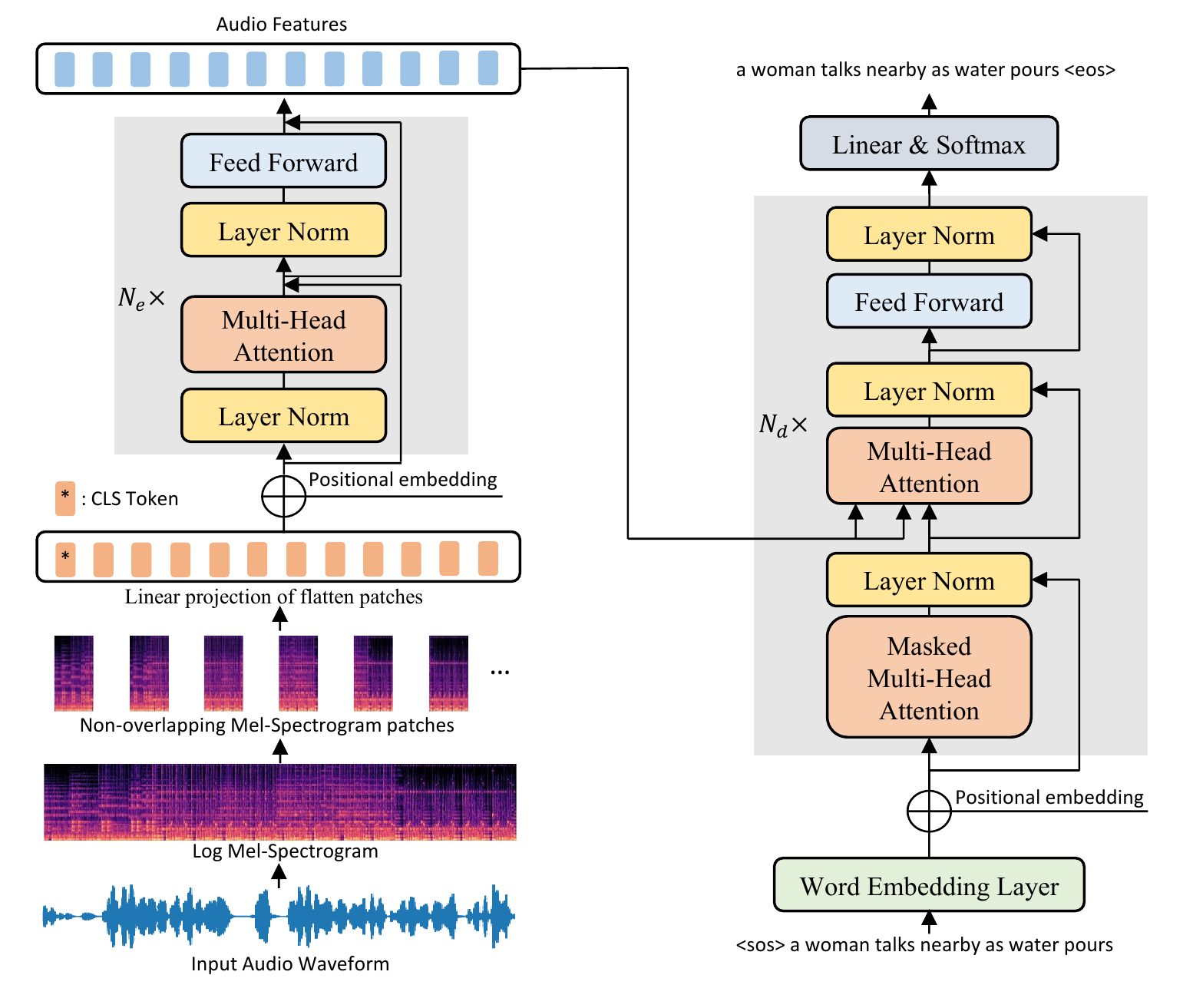}
  \caption{System overview of Audio Captioning Transformer, the encoder is on the left side while the decoder on the right side.}
  \label{fig:system_overview}
\end{figure}

\subsection{Encoder}
\label{ssec:encoder}
Let $X \in \mathbb R^{T\times F}$ denote the log mel-spectrogram of an audio clip, where $T$ is the number of time frames and $F$ is the number of mel bins. The log mel-spectrogram is first split into $N$ non-overlapping small patches $X_N = \{x_1,...,x_n\}$ along the time axis with size of $t \times F$ where $N=T/t$ and $t$ is the number of time frames of each patch. Then each mel-spectrogam patch is flattened to a 1D embedding and projected to a latent space through a learnable matrix $W_e \in \mathbb R^{(t\times F) \times d}$, where $d$ is the dimension of the latent embedding. In line with ViT and DeiT, a global learnable class token $
X_{{\rm cls}}\in \mathbb R^{1 \times d}$ is appended to the beginning of the patch sequences, which contains the global information for the audio clip. As the self-attention mechanism cannot capture position information \cite{vaswani2017attention}, a trainable positional embedding $X_{\rm pos}\in \mathbb R^{(T+1)\times d}$ is added to each patch embedding. Mathematically, the final input representation is given by
\begin{equation}
  \label{eqn:x_input}
  X_e = [X_{\rm cls} + W_eX] + X_{\rm pos}
\end{equation}

The ACT encoder consists of $N_{\rm e}$ stacked identical layers. Each layer contains two sub-layers, a multi-head self-attention layer and a position-wise fully-connected feed-forward layer. In the self-attention sub-layer, the input is first transformed into query $Q$, key $K$ and value $V$ through matrix multiplication with three learnable matrices $W_Q,W_K, W_V \in \mathbb R^{d \times d_k}$, where $d_k$ is the dimension of each attention head. Then the scaled dot-product attention is computed as 
\begin{equation}
  \label{eqn:attention}
  {\rm Attn}(Q,K,V) = {\rm Softmax}(\frac{QK^T} {\sqrt{d_k}})V
\end{equation}
Each self-attention layer contains $h$ attention heads which extends the model's ability to attend to different positions and creates multiple representation subspaces \cite{vaswani2017attention}. The outputs of heads are then aggregated through a linear transformation matrix $W_o \in \mathbb R^{(h\times d_k)\times d_k}$, which can be formulated as
\begin{equation}
  \label{eqn:multi_head}
  {\rm MultiHead}(Q,K,V) = {\rm Concat}({\rm head}_1, ..., {\rm head}_h)W_o
\end{equation}

The feed-forward network contains two linear layers with GLEU activation function and dropout applied between them. Layer normalization is applied before each sub-layer and a residual connection is employed around each of them, such that the output of each sub-layer is given by
\begin{equation}
  \label{eqn:sub_layer_pre_norm}
  X_{\rm out} = X_{\rm in} + {\rm Sub\_layer}({\rm LayerNorm}(X_{\rm in}))
\end{equation}
In order to make use of pre-trained models, the encoder architecture is the same as ViT and DeiT containing \num{12} encoder blocks and \num{12} heads with an embedding dimension of \num{768}.

\subsection{Decoder}
\label{ssec:decoder}
The ACT decoder contains three parts: a word embedding layer, a Transformer decoder block, and a linear layer. Each input word is embedded through the word embedding layer into a fixed dimension word vector and then fed into the Transformer decoder block. The word vectors are pre-trained by a Word2Vec model on all caption corpus \cite{mikolov2013efficient}.

The Transformer decoder consists of $N_{\rm d}$ identical stacked layers. There are two main differences compared to the ACT encoder block. First, the first self-attention sub-layer in the decoder is a masked self-attention because the caption generating process is causal and auto-regressive. Second, there is a new cross multi-head attention sub-layer between self-attention sub-layer and feed-forward sub-layer, which allows every position in the decoder to attend over all positions in the audio features extracted by the encoder \cite{vaswani2017attention}. The output of the decoder module is fed through a final linear layer with softmax activation function to output a probability distribution over the vocabulary. 

\begin{table}[!t]
    \centering
    \begin{tabular}{c c c c c}
    \hline 
    Model & embedding dim & \# layers ($N_{\rm d}$) & \# heads\\
    \hline 
    ACT\_s & 512 & 2 & 4 \\
    ACT\_m & 512 & 4 & 8 \\
    ACT\_l & 512 & 6 & 8 \\
    \hline
    \end{tabular}
    \caption{Variants of the proposed ACT decoder. }
    \label{table:models_arch}
\end{table}

The training objective of the model is to minimize the cross-entropy (CE) loss
\begin{equation}
  \label{eqn:ce_loss}
  \mathcal L_{\rm CE}(\theta)= - \frac{1}{T} \sum_{t=1}^T \log{p(y_t|y_{1:t-1},\theta)}
\end{equation}
where $y_t$ is the ground-truth word at time step $t$ and $\theta$ are the model parameters. The ``Teacher forcing'' strategy is used during training, i.e. each word to be predicted is conditioned on previous ground-truth words. We experiment with three models, which share the same encoder architecture described in Section \ref{ssec:decoder} but have different number of layers and heads in the decoder. Table \ref{table:models_arch} summarizes the parameters in the decoder of these models.

\section{Experiments}
\label{sec:exp}
\subsection{Dataset}
\label{ssec:dataset}

\subsubsection{AudioSet}
\label{sssec:audioset}
AudioSet is a large-scale audio dataset with an ontology of 527 sound classes \cite{audioset}. AudioSet contains more than \num{2} million 10-second audio clips extracted from YouTube videos. As some audio clips are no longer downloadable, there are \num{1934187} and \num{18887} audio clips in our training and evaluation set, respectively. Each audio clip can have one or more labels for their presented audio events.

\subsubsection{AudioCaps}
\label{ssec:audiocaps}
AudioCaps is the largest audio captioning dataset currently available with around 50k audio clips sourced from AudioSet \cite{kim2019audiocaps}. AudioCaps is divided into three splits. Each audio clip in the training set contains one human-annotated caption, while each contains five captions in the validation and test set. 

\subsection{Data pre-processing}
\label{ssec:dat_preprocess}
All audio clips in these two datasets are converted to 32k Hz and padded to 10-second long. Log mel-spectrograms extracted using a \num{1024}-points Hanning window with \num{50}\% overlap and \num{64} mel bins are used as the input features. Each log mel-spectrogram is split into \num{125} non-overlap small patches with the size of $64 \times 4$ along the time axis. SpecAugment \cite{park2019specaugment} is applied to augment the input features during training. 

Captions are tokenized and transformed to lower case with punctuation removed. To indicate the start and end of each caption, two special tokens ``\texttt{\textless sos\textgreater}'' and ``\texttt{\textless eos\textgreater}'' are padded. The vocabulary of AudioCaps contains \num{5277} distinct words.

\subsection{Audio tagging pre-training}
\label{ssec:pre_training}
As proved in previous works, Transformer requires more training data to achieve competitive performance with CNNs \cite{dosovitskiy2020image}. However, the amount of training data in audio processing area is much less than that in computer vision. Cross-modal transfer learning from ImageNet pre-trained models to audio-related tasks proves to be effective \cite{gong2021psla}. Thus we make use of pre-trained DeiT models for image classification to initialize the parameters in ACT encoder \cite{dosovitskiy2020image, touvron2021training}. As images have three channels and spectrograms just have one channel, we take the average of the weights from the patch embedding layer in DeiT in order to adapt it for spectrogram.

As pre-trained audio neural networks (PANNs) proved to perform well in audio captioning \cite{xinhao2021_t6}, we pre-train ACT encoder on AudioSet as an audio tagging task in order to solve the data scarcity problem and learn more generalized audio patterns. Audio tagging is a multi-classification task of predicting the presence or absence of sound classes within an audio clip \cite{kong2019weakly}. The class token output from the encoder is fed through a linear layer with sigmoid activation function to output the audio events probabilities. The model is trained to minimize the binary cross-entropy loss between the output of the model and the true label
\begin{equation}
  \label{eqn:at_loss}
  \mathcal L_{\rm BCE}(\theta) = - \sum_{n=1}^N (y_n \cdot \ln f(x_n) + (1-y_n)\cdot \ln (1-f(x_n))
\end{equation}
where $x_n$ is the $n$-th audio clip in AudioSet and $N$ is the number of training samples. $f(x_n) \in [0,1]^K$ is the output of the model and $y_n \in \{0,1\}^K$ is the true label where $K$ is the number of sound classes. The ACT encoder is pre-trained for \num{20} epochs with batch size of \num{128} and learning rate of \num{1e-4}, which achieves a mean average precision (mAP) of \num{0.43} on the evaluation set of AudioSet dataset. 
\begin{table*}[ht]
\centering
\begin{tabular}[\linewidth]{c c c c c c c c c c c} 
 \hline
 Model & BLEU$_{1}$ & BLEU$_{2}$ & BLEU$_{3}$ & BLEU$_{4}$ & ROUGE$_{L}$ & METERO & CIDEr & SPICE & SPIDEr \\ 
 \hline
 ACT\_s\_DeiT\_AudioSet & 0.643 & 0.483 & 0.352 & 0.249 & 0.469 & 0.218 & 0.669 & 0.160 & 0.415 \\
 ACT\_m\_DeiT\_AudioSet & 0.653 & \textbf{0.495} & \textbf{0.363} & \textbf{0.259} & \textbf{0.471} & 0.222 & 0.663 & 0.163 & 0.413 \\
 ACT\_l\_DeiT\_AudioSet & 0.647 & 0.488 & 0.356 & 0.252 & 0.468 & 0.222 & 0.679 & 0.160 & 0.420 \\
 \hline
 ACT\_m\_scratch & 0.567 & 0.411 & 0.285 & 0.191 & 0.417 & 0.187 & 0.501 & 0.127 & 0.314 \\
 ACT\_m\_DeiT & 0.606 & 0.445 & 0.319 & 0.224 & 0.445 & 0.207 & 0.586 & 0.147 & 0.367 \\
 \hline
 RNN+RNN \cite{kim2019audiocaps} & 0.614 & 0.446 & 0.317 & 0.219 & 0.450 & 0.203 & 0.593 & 0.144 & 0.369 \\
 CNN+RNN \cite{xu2021investigating} & \textbf{0.655} & 0.476 & 0.335 & 0.231 & 0.467 & \textbf{0.229} & 0.660 & \textbf{0.168} & 0.414 \\
 CNN+Transformer \cite{xinhao2021_t6} & 0.641 & 0.479 & 0.344 & 0.236 & 0.469 & 0.221 & \textbf{0.693} & 0.159 & \textbf{0.426} \\
 CNN+Transformer\_scratch \cite{xinhao2021_t6} & 0.610 & 0.461 & 0.334 & 0.234 & 0.455 & 0.206 & 0.629 & 0.144 & 0.386 \\
 \hline
\end{tabular}

\caption{Scores of the ACT model on the AudioCaps test set. DeiT: the ACT encoder is initialized with the parameters in DeiT, AudioSet: the ACT encoder is pre-trained on AudioSet.}
\label{table:tab_results} 
\end{table*}

\subsection{Experimental setups}
\label{ssec:exp_setup}
We train the proposed model for \num{30} epochs using Adam optimizer \cite{kingma2014adam} and a batch size of \num{32}. The learning rate is linearly increased to \num{1e-4} in the first five epochs using warm-up, which is then multiplied by \num{0.1} every \num{10} epochs. To mitigate over-fitting problem, dropout with rate of \num{0.2} is applied in the whole model. Label smoothing \cite{szegedy2016labelsmoothing} with a smoothing factor of \num{0.1} is used to avoid over-confident prediction. We use beam search with a beam size up to \num{5} to improve the decoding performance during inference stage.

\subsection{Evaluation metrics}
\label{ssec:metrics}

In line with previous works, we evaluate our methods using machine translation and captioning metrics \cite{tran2020wavetransformer}. BLEU$_{n}$, ROUGE$_{l}$ and METEOR are machine translation metrics. BLEU$_n$ is a modified precision metric with a sentence-brevity penalty, calculated as a weighted geometric mean over different length n-grams. ROUGE$_{l}$ calculates F-measures by counting the longest common subsequence. METEOR evaluates a caption by computing a harmonic mean of precision and recall based on explicit word-to-word matches between the caption and given references. Captioning metrics contain CIDE$_{r}$, SPICE and SPIDE$_r$. CIDE$_{r}$ calculates the cosine similarity between term frequency inverse document frequency (TF-IDF) weighted n-grams. SPICE creates scene graphs for captions and calculates F-score based on tuples in the scene graphs. SPIDE${r}$ is the average of SPICE and CIDE$_{r}$ and is selected as the official ranking metric in DCASE challenge, the SPICE score ensures captions are semantically faithful to the audio content, while CIDE$_{r}$ score ensures captions are syntactically fluent. 
\section{Results}
\label{sec:results}
\subsection{Performance comparison}
\label{ssec:comparison}
Table \ref{table:tab_results} presents the results on AudioCaps test set. We compare the proposed ACT model with three representative audio captioning models, ``RNN+RNN'' \cite{kim2019audiocaps}, ``CNN+RNN'' \cite{xu2021investigating} and ``CNN+Transformer'' \cite{xinhao2021_t6}. In these models, CNNs are all pre-trained on upstream audio-related tasks. As can be seen in Table \ref{table:tab_results} that the ACT model outperforms ``RNN+RNN'' model substantially in all evaluation metrics and achieves slightly higher scores than ``CNN+RNN'' model in most metrics. Compared with the state-of-the-art ``CNN+Transformer'' approach, ACT model outperforms it in machine translation metrics but gives slightly lower scores in CIDE$_r$. As machine translation metrics are based mostly on n-grams, these results show that the ACT model has better ability in generating words accurately. In addition, training an ACT model is faster than ``CNN+Transformer'' architecture, where the former takes less than five minutes for one epoch and ``CNN+Transformer'' needs seven minutes in our experiments. In summary, the ACT model shows competitive performance as compared to other state-of-the-art approaches, and it is simple as it based only on the self-attention mechanism.

\subsection{Ablation studies}
\label{ssec:ablation_studies}
The ablation studies are carried out to investigate the effectiveness of the pre-trained encoder and the influence of the hyper-parameters in the decoder. From the experimental results, we can see that pre-training the ACT encoder can boost the performance significantly. Even only using the pre-trained DeiT model, which is originally trained for image classification task, can bring significant performance gains in all the evaluation metrics. Pre-training on AudioSet as an audio tagging task further improves the system to approach the state-of-the-art performance. We also compare the ACT model with the ``CNN+Transformer'' model both trained from scratch, the results show that the ACT model performs worse than ``CNN+Transformer'' without encoder pre-training. These results suggest that pre-training the ACT encoder with a large dataset is important, and prove that Transformer network needs more training data than CNNs to achieve competitive performance. 

We perform experiments on the three models with different numbers of layers and heads in the decoder. From the observations, the ACT model is slightly sensitive to the choice of hyper-parameters in the decoder. These three models achieve similar performance, among which ACT\_m with four decoder layers performs better in machine translation metrics, while ACT\_l achieves higher CIDE$_r$ and SPIDE$_r$ scores. The ACT model only needs shallow Transformer decoder layers compared to machine translation models in natural language tasks which typically contain \num{12} Transformer decoder layers \cite{vaswani2017attention}. There might be two reasons. First, the amount of training data in audio captioning is far less than data in natural language processing tasks. Second, the length of the audio captions are usually shorter than sentences in the natural language tasks.
\section{Conclusion}
\label{sec:conclusion}
We have presented a novel audio captioning model, Audio Captioning Transformer (ACT), which is a full Transformer model based on the self-attention mechanism. The encoder of the proposed ACT model can model the global and fine-grained information within an audio signal simultaneously, and has better ability to capture temporal relationships between audio events than CNNs. Experimental results show that the ACT model can outperform other state-of-the-art audio captioning systems in most metrics. Further research should be carried out to adapt the ACT model for audio clips of varied lengths.

\section{ACKNOWLEDGMENT}
\label{sec:ackn}
This work is partly supported by grant EP/T019751/1 from the Engineering and Physical Sciences Research Council (EPSRC), a Newton Institutional Links Award from the British Council, titled ``Automated Captioning of Image and Audio for Visually and Hearing Impaired" (Grant number 623805725) and a Research Scholarship from the China Scholarship Council (CSC) No. 202006470010. 

\bibliographystyle{IEEEtranN}
\bibliography{refs}

\end{sloppy}
\end{document}